\documentclass[10pt,aps,prx,twocolumn,notitlepage,showpacs,superscriptaddress,nofootinbib]{revtex4-1}
\usepackage[bottom]{footmisc}
\usepackage{amsthm,amsmath,amsfonts,amssymb,verbatim,color}
\usepackage{graphicx}
\usepackage{bm}
\usepackage{epsfig,slashed}
\usepackage{amssymb}
\usepackage{amsfonts}
\usepackage{tikz}
\usepackage{xparse}
\usepackage{ifthen}
\usepackage{lipsum}
\usepackage[colorlinks=true,citecolor=blue,
linkcolor=blue,urlcolor=blue]{hyperref}
\usepackage[caption=false]{subfig}
\begin{document}

\newcommand{\tr}{\mathop{\mathrm{tr}}}
\newcommand{\bsigma}{\boldsymbol{\sigma}}
\newcommand{\re}{\mathop{\mathrm{Re}}}
\newcommand{\im}{\mathop{\mathrm{Im}}}
\renewcommand{\b}[1]{{\boldsymbol{#1}}}
\newcommand{\diag}{\mathrm{diag}}
\newcommand{\sign}{\mathrm{sign}}
\newcommand{\sgn}{\mathop{\mathrm{sgn}}}

\newcommand{\cl}{\mathrm{cl}}
\newcommand{\mb}{\bm}
\newcommand{\ua}{\uparrow}
\newcommand{\da}{\downarrow}
\newcommand{\ra}{\rightarrow}
\newcommand{\la}{\leftarrow}
\newcommand{\mc}{\mathcal}
\newcommand{\bs}{\boldsymbol}
\newcommand{\lra}{\leftrightarrow}
\newcommand{\nn}{\nonumber}
\newcommand{\half}{{\textstyle{\frac{1}{2}}}}
\newcommand{\mf}{\mathfrak}
\newcommand{\MF}{\text{MF}}
\newcommand{\IR}{\text{IR}}
\newcommand{\UV}{\text{UV}}
\newcommand{\sech}{\mathrm{sech}}

\title{Transition between algebraic and $\mathbb{Z}_2$ quantum spin liquids at large $N$}

\author{Rufus Boyack}
\affiliation{Department of Physics, University of Alberta, Edmonton, Alberta T6G 2E1, Canada}
\affiliation{Theoretical Physics Institute, University of Alberta, Edmonton, Alberta T6G 2E1, Canada}

\author{Chien-Hung Lin}
\affiliation{School of Physics and Astronomy, University of Minnesota, Minneapolis, Minnesota 55455, USA}

\author{Nikolai Zerf}
\affiliation{Institut f\"ur Physik, Humboldt-Universit\"at zu Berlin, Newtonstra{\ss}e 15, D-12489 Berlin, Germany}

\author{Ahmed Rayyan}
\affiliation{Department of Physics, University of Alberta, Edmonton, Alberta T6G 2E1, Canada}

\author{Joseph Maciejko}
\affiliation{Department of Physics, University of Alberta, Edmonton, Alberta T6G 2E1, Canada}
\affiliation{Theoretical Physics Institute, University of Alberta, Edmonton, Alberta T6G 2E1, Canada}
\affiliation{Canadian Institute for Advanced Research, Toronto, Ontario M5G 1Z8, Canada}

\date\today

\begin{abstract}
We present a field theory description of a quantum phase transition in two spatial dimensions between a $U(1)$ algebraic spin liquid with $N$ flavors of gapless two-component Dirac fermionic spinons and a gapped $\mathbb{Z}_2$ spin liquid. This transition is driven by spinon pairing and concomitant Higgsing of the emergent $U(1)$ gauge field. For sufficiently large $N$ we find a quantum critical point with non-Gaussian exponents that is stable against instanton proliferation. We compute critical exponents using either $1/N$ or 
$\epsilon$ expansions, and give estimates of the critical value of $N$ below which the quantum critical point disappears.
\end{abstract}

\maketitle

\section{Introduction}

Quantum spin liquids are a fascinating class of exotic quantum states of matter, characterized by long-range entanglement, fractionalized excitations, and topological/quantum order~\cite{QSLreview}. Originally proposed by Anderson~\cite{anderson1973}, quantum spin liquid ground states are believed to arise in frustrated quantum antiferromagnets in which quantum fluctuations prevent the formation of long-range magnetic order even at zero temperature. These phases of matter have long eluded experimental detection, until the identification in recent years of several promising candidate materials, including the organic salts $\kappa$-(BEDT-TTF)$_2$Cu$_2$(CN)$_3$~\cite{shimizu2003} and EtMe$_3$Sb[Pd(dmit)$_2$]$_2$~\cite{itou2008}, ZnCu$_3$(OH)$_6$Cl$_2$ (herbertsmithite)~\cite{han2012,fu2015}, and the Kitaev material $\alpha$-RuCl$_3$~\cite{baek2017,do2017}.

On the theoretical side, a variety of methods have been developed to characterize and classify possible spin liquid ground states, including trial wavefunctions (such as Anderson's celebrated resonating valence bond wavefunction~\cite{anderson1973,anderson1987,baskaran1987}), quantum dimer models~\cite{rokhsar1988,moessner2001}, exactly soluble models~\cite{kitaev2003,kitaev2006}, and large-$N$ methods~\cite{affleck1988,read1991}, not to mention numerical methods such as exact diagonalization and the density-matrix renormalization group. In particular, large-$N$ methods---whereby the usual $SU(2)$ spin symmetry is enlarged to $SU(N)$ or $Sp(N)$, and $1/N$ is treated as a small parameter---lead to a description of spin liquids in terms of parton gauge theories, in which electrically neutral spin-$1/2$ fermionic or bosonic spinons interact with an emergent gauge field. In two spatial dimensions, and limiting ourselves to theories with fermionic spinons, such theories predict two broad classes of spin liquids: $U(1)$ spin liquids, where gapless spinons forming Dirac points~\cite{affleck1988} or a Fermi surface~\cite{baskaran1987,lee2005} interact with a $U(1)$ gauge field, and $\mathbb{Z}_2$ spin liquids, where the gauge structure is broken down to $\mathbb{Z}_2$~\cite{wen1991}, by spinon pairing for instance.

An important question in the parton gauge theory description of spin liquids is whether a given proposed spin liquid state, understood as spinons forming a particular emergent bandstructure, is stable against gauge fluctuations. In the $\mathbb{Z}_2$ case, gauge fluctuations are gapped and thus can only mediate short-range interactions between spinons, which are irrelevant in the low-energy limit. In the $U(1)$ case the gauge fluctuations are gapless, and while some approaches~\cite{lee2008,metlitski010} indicate that a spinon Fermi surface is stable against such gauge fluctuations, this issue remains under debate in light of recent work~\cite{Holder_2015} suggesting the corresponding field theory is nonrenormalizable. A system with $N$ flavors of Dirac spinons is stable against gauge fluctuations for sufficiently large $N$~\cite{kim1999,rantner2002,hermele2004} and is known as an algebraic spin liquid. In particular, it is stable against instanton proliferation in the large-$N$ limit~\cite{hermele2004}, which in the absence of sufficient amounts of gapless matter drives spinon confinement~\cite{polyakov1975,polyakov1977}. As its name suggests, the algebraic spin liquid exhibits power-law correlations in the low-energy, long-wavelength limit, and is described in that limit by an interacting (2+1)-dimensional conformal field theory.

While the spin liquid states described above are relatively well understood theoretically, much less is known about phase transitions between such states. In two spatial dimensions spin liquids are only sharply defined in the limit of zero temperature, and we thus focus our attention on quantum phase transitions. In the parton approach such transitions, if continuous, are naturally described by critical points of interacting gauge theories, and fall outside the usual Landau-Ginzburg-Wilson paradigm in which the order parameter field is the only fluctuating degree of freedom. Much attention has been devoted to exotic transitions between ordered phases of quantum antiferromagnets---such as the N\'eel to valence-bond-solid transition---in which emergent gauge fields are confined in the ordered phases but undergo deconfinement exactly at the critical point~\cite{senthil2004,senthil2004b}. Although important progress has been made in the understanding of certain transitions between symmetry-protected topological phases~\cite{grover2013,lu2014}, which can be thought of as short-range entangled analogs of spin liquids without deconfined emergent gauge fields, there have been relatively few studies of transitions between long-range entangled spin liquid phases. The theory of anyon condensation~\cite{bais2009} provides a formal classification of allowed transitions between topologically-ordered spin liquids. In the context of parton gauge theories, which is our focus here, we note studies of transitions between chiral and $\mathbb{Z}_2$ spin liquids~\cite{barkeshli2013}, between $U(1)$ spin liquids with a spinon Fermi surface and $\mathbb{Z}_2$ spin liquids~\cite{metlitski2015}, and between algebraic and chiral spin liquids~\cite{he2015,janssen2017}.

In this paper we study a model of a phase transition between an algebraic spin liquid with $N$ flavors of emergent Dirac fermions and a gapped, topologically-ordered $\mathbb{Z}_2$ spin liquid, driven by fermion pairing. Making use of both large-$N$ and $\epsilon$-expansion techniques, we argue that this transition is continuous for sufficiently large $N$ and is stable in that limit against instanton proliferation. Both expansions suggest there exists a critical number $N_c$ of fermion flavors below which the quantum critical point found becomes unstable. At leading order, we find $N_c^\text{inst}=11.32$ in the large-$N$ expansion and $N_c=34.90$ in the $\epsilon$ expansion. However, we find that next-to-leading order ($\epsilon^2$) corrections tend to significantly reduce the latter value. Of course, nonperturbative approaches such as numerical methods are required to definitively settle this issue.

The paper is structured as follows. In Sec.~\ref{sec:model} the model is presented, and then in Sec.~\ref{sec:largeN} we give its exact solution in the large-$N$ limit in fixed spacetime dimension $d=3$. We determine the correlation length exponent $\nu$ in that limit and estimate that the critical point found is stable against instanton proliferation for $N>11.32$. In Sec.~\ref{sec:epsilon} we complement the large-$N$ analysis by a finite-$N$ renormalization group (RG) analysis in $d=4-\epsilon$ spacetime dimensions. A fixed point is found at one-loop order for $N>34.90$ and shown to correspond to the critical point found in Sec.~\ref{sec:largeN}; $\nu$ as well as the scaling dimension of a fermion bilinear operator are computed to leading order in $\epsilon$. We then derive and use three-loop RG flow equations to estimate corrections to $N_c$, which we find to be negative when extrapolated to physical dimensions. We briefly conclude and suggest directions for future research in Sec.~\ref{sec:conclusion}.

\section{Model}
\label{sec:model}

The model is a (2+1)D quantum field theory of $N$ two-component Dirac fermions $\psi_a$, $a=1,2,\ldots,N$, interacting with a dynamical $U(1)$ gauge field $A_\mu$ and a complex scalar field $\phi$, with imaginary-time Lagrangian
\begin{align}\label{L}
\mathcal{L}=&\sum_{a=1}^{N}i\bar{\psi}_a(\slashed{\partial}+ie\slashed{A})\psi_a+\frac{1}{4}F_{\mu\nu}^2+|(\partial_\mu+2ieA_\mu)\phi|^2
\nn\\{}&+m^2|\phi|^2+\lambda^2|\phi|^4+h\sum_{a=1}^{N}\left(\phi^*\psi_a^Ti\sigma_2\psi_a+\text{H.c.}\right).
\end{align}
The usual Pauli spin matrices are denoted by $\sigma_1,\sigma_2,\sigma_3$. Such a model would arise from the parton description of a microscopic spin model on a 2D lattice with $SU(k)$ symmetry, with $N$ a multiple of $k$ (see, e.g., Ref.~\cite{WenBook}). The precise relation between $N$ and $k$ depends on the particular model; for instance, for the proposed algebraic spin liquid ground state of the ($k=2$) Heisenberg model on the kagom\'e lattice~\cite{ran2007,hermele2008} one has $N=2k$. We define the Dirac conjugate $\bar{\psi}_a=-i\psi^\dag_a\gamma_0$, the Euclidean Dirac matrices $\gamma_0=\sigma_3$, $\gamma_1=\sigma_1$, $\gamma_2=\sigma_2$, and use Feynman's slashed notation $\slashed{X}\equiv\gamma_\mu X_\mu$ with $\mu=0,1,2$ a spacetime index. The field-strength tensor for the emergent gauge field is $F_{\mu\nu}=\partial_\mu A_\nu-\partial_\nu A_\mu$, $e$ is the gauge charge, and $m$, $\lambda$, and $h$ are phenomenological couplings. The theory has an $O(N)$ global flavor symmetry under $\psi_a\rightarrow W_{ab}\psi_b$, $\bar{\psi}_a\rightarrow \bar{\psi}_b W_{ba}^T$ with $W$ an arbitrary $N\times N$ orthogonal matrix, as well as a $U(1)$ gauge symmetry under $\psi_a(x)\rightarrow e^{i\theta(x)}\psi_a(x)$, $\bar{\psi}_a(x)\rightarrow e^{-i\theta(x)}\bar{\psi}_a(x)$, $\phi(x)\rightarrow e^{2i\theta(x)}\phi(x)$.

The complex scalar field $\phi$ is the order parameter for the transition between the algebraic spin liquid and the $\mathbb{Z}_2$ spin liquid, which can be seen as follows. The tuning parameter for the transition is the scalar mass squared parameter $m^2$. For $m^2>0$, the scalar is gapped and can be integrated out perturbatively, yielding only finite corrections to the Maxwell term for the gauge field as well as generating interactions such as contact four-Fermi terms, which are irrelevant in the large-$N$ limit~\cite{Xu_2008}. Only the first two terms in Eq.~(\ref{L}) matter for the low-energy behavior, and one recovers the theory of the algebraic spin liquid with $N$ Dirac fermion flavors~\cite{affleck1988,kim1999,rantner2002}. For $m^2<0$, one obtains a Mexican hat potential for the scalar, and the $U(1)$ gauge field is Higgsed to $\mathbb{Z}_2$. In the mean-field sense, $\phi$ and the fermion bilinear $\sum_a\psi_a^Ti\sigma_2\psi_a$ develop an expectation value, i.e., the fermions form $s$-wave Cooper pairs and are gapped out. This corresponds to a fully gapped $\mathbb{Z}_2$ spin liquid. The massless theory $m^2=0$ thus describes a quantum phase transition between $U(1)$ and $\mathbb{Z}_2$ spin liquids.

A few comments are in order. First, the chemical potential is set to zero, i.e., at the Dirac point, otherwise we would be describing a spin liquid with a spinon Fermi surface. Second, there are no fermion mass terms because of time-reversal, parity, and translation symmetries~\cite{wen2002}. Third, one might object to the use of a relativistic kinetic term for the scalar, since the original microscopic lattice model is nonrelativistic. However, if one imagines that pairing originates from residual interactions between the spinons, then $\phi$ should be viewed as a spinon pair, and as such its low-energy dynamics will be generated by integrating out high-energy Dirac fermion modes and will inherit the relativistic character of the latter. Fourth, even in this case, one should consider a Lagrangian with different velocities for the fermion, scalar, and gauge field. However, it was recently shown that in (2+1)D gauge theories with fermions and scalars and Yukawa-type couplings between the latter, all three velocities flow to a common value in the infrared limit~\cite{roy2016b}. The Lagrangian in Eq.~(\ref{L}) should thus be understood as the infrared effective Lagrangian for the $U(1)$-$\mathbb{Z}_2$ spin liquid transition. Finally, we point out a related study by Wen~\cite{wen2002}, in which it was shown with the help of a theory similar to ours, but without a Yukawa coupling nor tunable mass term for the scalar, that the algebraic spin liquid is perturbatively stable against the Higgs mechanism. This establishes that the algebraic and $\mathbb{Z}_2$ spin liquids are distinct stable phases of matter, but does not address the nature of a phase transition between the two, which is necessarily a nonperturbative phenomenon.

\section{Large-$N$ limit}
\label{sec:largeN}

To make the arguments in the preceding section more rigorous, and to convincingly show the existence of a stable critical point in a controlled limit, we study the model described by Eq.~(\ref{L}) in the large-$N$ limit.
In the next two subsections we will study the saddle-point condition along with fluctuations. We then discuss the issues of compactness of the gauge group and instantons. Finally, by considering corrections to scaling we obtain the critical couplings for this theory for the purpose of comparing with the results of the $\epsilon$ expansion in Sec.~\ref{sec:epsilon}.

\subsection{Saddle point}
\label{sec:SaddlePoint}

In the large-$N$ limit the coupling constants are expected to scale as $h,\lambda,e\propto1/\sqrt{N}$, 
therefore we scale these quantities by $1/\sqrt{N}$ in the Lagrangian in Eq.~(\ref{L}). 
Furthermore, in arbitrary $d$-dimensional spacetime these couplings are dimensionful quantities; by introducing a large momentum cutoff $\Lambda$ they can be expressed in terms of dimensionless variables as follows:
$h\rightarrow\Lambda^{2-d/2}h,\lambda\rightarrow\Lambda^{2-d/2}\lambda,e\rightarrow\Lambda^{2-d/2}e$.
After rescaling the fields by $\phi\rightarrow\phi\sqrt{N}/\left(h\Lambda^{2-d/2}\right)$ and $A_{\mu}\rightarrow\sqrt{N}A_{\mu}/\Lambda^{2-d/2}$, 
one finally obtains the following Lagrangian:
\begin{eqnarray}
\mathcal{L} & = & \sum_{a=1}^{N}i\bar{\psi}_{a}\left(\slashed{\partial}+ie\slashed{A}\right)\psi_{a}+N\Lambda^{d-4}\frac{1}{h^{2}}\left|\left(\partial_{\mu}+2ieA_{\mu}\right)\phi\right|^{2}\nonumber\\
&  & +N\Lambda^{d-4}\frac{m^{2}}{h^{2}}\left|\phi\right|^{2}+N\Lambda^{d-4}\frac{\lambda^{2}}{h^{4}}\left|\phi\right|^{4}\nonumber \\
&  & +\frac{1}{4}N\Lambda^{d-4}F_{\mu\nu}^{2} +\sum_{a=1}^{N}\left(\phi^{*}\psi_{a}^{T}i\sigma_{2}\psi_{a}+\mathrm{H.c.}\right).
\end{eqnarray}
In this section we work in the Landau gauge, introducing a gauge-fixing term in the Lagrangian $\sim\left(\partial_{\mu}A_{\mu}\right)^2/\left(2\xi\right)$ and setting $\xi=0$ at the end. 
We introduce the four-component Nambu spinor $\Psi_a=\left(\psi_a,i\sigma_{2}\bar{\psi}^{T}_a\right)^{T}$, which allows the first and last terms in the above Lagrangian to be combined into $\sum_{a=1}^{N}\frac{1}{2}\Psi_{a}^{T}\mathcal{C}\mathcal{G}^{-1}\Psi_{a}$ with $\mathcal{C}=i\sigma_{2}\oplus i\sigma_{2}$~\cite{zerf2016}. Here the inverse Nambu Green's function is $\mathcal{G}^{-1}=\mathcal{G}_{0}^{-1}+X_{A}+X_{\phi}$,
and the matrices (in Nambu space) are
\begin{eqnarray}
X_{\phi}=\left(\begin{array}{cc}
2\phi^{*} & 0\\
0 & -2\phi
\end{array}\right), && X_{A}=\left(\begin{array}{cc}
0 & e\slashed{A}\\
-e\slashed{A} & 0
\end{array}\right), \nonumber
\\
\mathcal{G}_{0}^{-1}&=&\left(\begin{array}{cc}
0 & i\slashed{\partial}\\
i\slashed{\partial} & 0
\end{array}\right).
\end{eqnarray}

The fermions can then be integrated out, using the functional integral formula for Majorana fermions. Following the analysis for the Gross-Neveu-Yukawa model~\cite{moshe2003}, we integrate out only $\left(N-1\right)$ fermions to obtain the following fermionic contribution to the action:
$S_{f}=-\frac{1}{2}\left(N-1\right)\mathrm{Tr\,ln}\left(\mathcal{G}^{-1}\right)$. 
The remaining fermion is labeled $\Psi$, and the full action then becomes 
\begin{eqnarray}\label{FullActionLargeN}
S\left[\Psi,\phi,A_{\mu}\right] & = & \int d^{d}x\biggl[\frac{1}{2}\Psi^{T}\mathcal{C}\mathcal{G}^{-1}\Psi + \frac{1}{4}N\Lambda^{d-4}F_{\mu\nu}^{2}\nonumber\\
&  & +N\Lambda^{d-4}\frac{1}{h^{2}}\left|\left(\partial_{\mu}+2ieA_{\mu}\right)\phi\right|^{2}\nonumber\\
&  & +N\Lambda^{d-4}\frac{m^{2}}{h^{2}}\left|\phi\right|^{2}+N\Lambda^{d-4}\frac{\lambda^{2}}{h^{4}}\left|\phi\right|^{4}\biggr]\nonumber\\
&  & -\frac{1}{2}\left(N-1\right)\mathrm{Tr\,ln}\left(\mathcal{G}^{-1}\right).
\end{eqnarray}
In the large-$N$ limit, since the rescaled $e$ is assumed to be $\mathcal{O}\left(N^{0}\right)$, the first term in the action can be neglected. Thus, the large-$N$ action reduces to
\begin{eqnarray}
\label{LargeNAction}
&&S\left(N\gg1\right) \rightarrow N\biggl[\Lambda^{d-4}\int d^{d}x\biggl\{\frac{1}{h^{2}}\left|\left(\partial_{\mu}+2ieA_{\mu}\right)\phi\right|^{2}\nonumber\\
&  & +\frac{m^{2}}{h^{2}}\left|\phi\right|^{2}+\frac{\lambda^{2}}{h^{4}}\left|\phi\right|^{4}+\frac{1}{4}F_{\mu\nu}^{2}\biggr\} -\frac{1}{2}\mathrm{Tr\,ln}\left(\mathcal{G}^{-1}\right)\biggr].
\end{eqnarray}

In the $N\rightarrow\infty$ limit, the saddle-point approximation to the functional integral is exact. We thus set the first variation of the large-$N$ action in Eq.~(\ref{LargeNAction}), with respect to the fields $\phi(y)$ and $A_{\mu}(y)$, equal to zero: $ \delta S/\delta\phi(y)=0, \delta S/\delta A_{\mu}(y)=0$. The solution to these ``gap equations" are the saddle-point, or classical, field configurations $\phi_{\cl}(y)$ and $A_{\mu,\cl}(y)$. For the complex scalar field, we look for a uniform configuration $\phi_{\cl}(y)=M$. This saddle-point solution for $\phi$ of course spontaneously breaks the global $U(1)$ symmetry, and so it is only for the case where $M=0$ that the gauge-field propagator (see Sec.~\ref{sec:Fluctuations}) will be gauge invariant. In three (Euclidean) dimensions the only possible translationally invariant nonzero solution for the gauge field corresponds to a uniform magnetic field $\nabla\times\boldsymbol{A}_{\cl}=\boldsymbol{B}_{\cl}\neq 0$, which breaks Lorentz symmetry. We will restrict ourselves to Lorentz invariant solutions $\boldsymbol{B}_{\cl}=0$, for which one can take $A_{\mu,\cl}=0$. We now show that this ansatz is consistent with the saddle-point condition for the gauge field. Setting $\delta S/\delta A_{\mu}(y)=0$, and then inserting $\phi_{\cl}(y)=M$ and $A_{\mu,\cl}(y)=0$, gives the following condition: $0=e\mathrm{Tr}\mathcal{G}(y,y)[i\sigma_{2}\otimes\gamma_{\mu}]$; from the structure of the Nambu Green's function, and since the trace of the Pauli matrices is zero, it follows that the right-hand side of this expression indeed vanishes.

To obtain the gap equation for the scalar field, we set $\delta S/\delta \phi(y)=0$, and then insert $\phi_{\cl}(y)=M$ and $A_{\mu,\cl}(y)=0$. The trivial solution to this is $M=0$, while the non-trivial condition for $\phi_{\cl}=M$ can be written as 
\begin{equation}
\frac{m^{2}}{h^{2}}+\frac{2\lambda^{2}}{h^{4}}\left|M\right|^{2} = 4\Lambda^{4-d}\int_{|q|<\Lambda}\frac{d^{d}q}{\left(2\pi\right)^{d}}\frac{1}{q^{2}+4\left|M\right|^{2}}.
\end{equation}
The integral appearing on the right hand side can be computed explicitly in the limit $\Lambda\rightarrow\infty$, for $2<d<4$. In $d=3$, one obtains $-2|M|/(4\pi)$. 

At the quantum critical point, the mass is $m^{2}=m_{\mathrm{c}}^{2}$ and the order parameter vanishes: $M=0$. If we define $r= \Lambda^{d-4}\left(m^{2}-m_{\mathrm{c}}^{2}\right)/h^{2}$ and 
insert this into the original gap equation, then for finite $\Lambda$ with $d=3$, we obtain 
\begin{equation}
r+\frac{2\lambda^{2}}{h^{4}\Lambda}\left|M\right|^{2}+\frac{2\left|M\right|}{\pi}=0.
\end{equation}
Near the quantum critical point $\left|M\right|^{2}\ll\left|M\right|$ so that $\left|M\right|=\pi\left(-r\right)/2$, where in the broken symmetry phase $r<0$.

Since $M$ is the (spontaneously generated) fermion mass in the broken symmetry phase, the spectrum acquires a gap $\Delta= 2|M|\propto (m^{2}_{c}-m^2)$.
Near a quantum phase transition, the gap scales as $\Delta\propto J|u-u_{c}|^{z\nu}$, where $J$ is some microscopic energy scale, $u$ is the coupling that tunes the transition (coefficient of the relevant operator at the fixed point), and $z$ and $\nu$ are the dynamic critical exponent and correlation length exponent, respectively~\cite{QPT}. For a Lorentz-invariant critical point (as here), one has $z=1$. Since $m^{2}\propto u$ is the parameter that tunes the transition, we obtain
\begin{equation}
\nu = 1 + \mathcal{O}\left(1/N\right), 
\end{equation}
as in the pure (ungauged) Gross-Neveu-Yukawa model~\cite{moshe2003}. This is a non-Gaussian exponent, and the critical point is strongly coupled already in the $N\rightarrow\infty$ limit.

\subsection{Fluctuations}
\label{sec:Fluctuations}

One can also calculate the fermion, scalar, and gauge-field propagators in the large-$N$ limit by considering fluctuations about the classical field configurations: we set $\phi(x)=\phi_{\cl}+\sigma(x)/\sqrt{N}, A_{\mu}(x)=A_{\mu,\cl} + B_{\mu}(x)/\sqrt{N}$. 
Inserting these into the trace-logarithm formula and then expanding about the classical field configurations, we observe that the terms quadratic in fluctuations are $\mathcal{O}(1/N^0)$ while higher order terms are $\mathcal{O}(1/\sqrt{N})$ and disappear in the $N\rightarrow\infty$ limit. 
Thus, the quadratic expansion of the logarithm is exact in the large-$N$ limit. 

At the critical point, where $M=0$, the fluctuation contribution to the action can be written as
\begin{align}
\int\frac{d^dq}{(2\pi)^d}\biggl[\frac{1}{2}B_{\mu}(q)K_{\mu\nu}^{B}(q)B_{\nu}(-q)+\sigma^{*}(q)K_{\sigma^{*},\sigma}(q)\sigma(q)\biggr],
\end{align}
where, in $d=3$ dimensions, one obtains
\begin{equation}\label{eq:B_Response}
K^{B}_{\mu\nu}\left(q\right)=\frac{1}{16}e^{2}\left(\delta_{\mu\nu}-\frac{q_{\mu}q_{\nu}}{q^{2}}\right)\left|q\right|.
\end{equation}
As required, this (current-current) correlation function is gauge invariant: $q_{\mu}K_{\mu\nu}^{B}(q)=0$. Thus the large-$N$ action for the gauge field is the same as in pure massless (2+1)D quantum electrodynamics ($\mathrm{QED}_{3}$)~\cite{Bhattacharyya_1987}, i.e., it corresponds to the algebraic spin liquid fixed point. (For the case of four-component spinors the prefactor would be 1/8, in agreement with Ref.~\cite{kim1999}.) Similarly, for the scalar field, one obtains
 \begin{equation}\label{eq:Sigma_Response}
K_{\sigma^{*},\sigma}\left(q\right)=\frac{1}{4}\left|q\right|.
\end{equation}
In principle, one can also look at fluctuations of the (remaining) fermion field $\Psi$ in Eq.~(\ref{FullActionLargeN}), as done for the Gross-Neveu-Yukawa model~\cite{moshe2003}; at criticality, the fermion propagator simply assumes its free form, $1/\slashed{p}$. In the Gross-Neveu-Yukawa model, one can then read off the scaling dimensions of the scalar and fermion fields at criticality in the large-$N$ limit; here, however, those scaling dimensions are not gauge-invariant quantities.

We have thus found that in the large-$N$ limit the effective action at the critical point reduces to decoupled copies of QED$_3$ and the ungauged Gross-Neveu-Yukawa model (with complex scalars). However, higher-order corrections in the $1/N$ expansion---which we defer to future work---will involve coupling between the gauge and matter sectors, and lead to deviations from the pure QED$_3$ or Gross-Neveu-Yukawa exponents.

\subsection{Instantons}

We now turn to the issue of instantons. The analysis just performed shows that, in the large-$N$ limit, there exists a stable conformal field theory in 2+1 dimensions describing a continuous transition between an algebraic spin liquid and a gapped $\mathbb{Z}_2$ spin liquid. However, the analysis has neglected the compactness of the $U(1)$ gauge group, which follows from the fact that the microscopic theory is originally defined on a lattice. A compact $U(1)$ gauge theory on a lattice admits instanton events corresponding to the insertion of $2\pi$ emergent magnetic flux in a localized region of spacetime~\cite{polyakov1975,polyakov1977}. To show that the transition is truly described by the conformal field theory found above, we need to show that monopole operators, which perform the above flux insertions, are irrelevant at the transition.

To show that monopole operators are irrelevant, one must calculate the scaling dimension of the most relevant/least irrelevant symmetry-preserving monopole operator at the fixed point and show that it exceeds the spacetime dimension $d=3$. At the algebraic spin liquid fixed point, the scaling dimension of the least irrelevant monopole operator (corresponding to $2\pi$ flux insertion) was shown in Ref.~\cite{hermele2004} to be proportional to $N$ in the large-$N$ limit, and explicitly calculated in the large-$N$ limit in Ref.~\cite{borokhov2002}, with $1/N$ corrections included in Ref.~\cite{pufu2014}:
\begin{align}\label{MonopoleQED3}
\Delta_{2\pi}^{\text{QED}_3}=0.265N-0.0383+\mathcal{O}(1/N).
\end{align}
At this order in the $1/N$ expansion, this implies that the $2\pi$ monopole operator is irrelevant for $N>11.47$~\cite{pufu2014}. Whether this operator is symmetry-preserving or not depends on the microscopic spin model, but we will not discuss such considerations here and conservatively assume that the smallest monopole operator (and therefore also the least irrelevant~\cite{borokhov2002}) preserves all the symmetries.

We now turn to the scaling dimension of monopole operators at the quantum critical point. According to the state-operator correspondence of conformal field theory~\cite{rychkov2016,simmons-duffin2016}, the scaling dimension of a monopole operator with $2\pi n$ flux, $n\in\mathbb{Z}$, is equal to the ground-state energy of the conformal field theory quantized on the unit sphere $S^2$ with $2\pi n$ magnetic flux through the sphere (i.e., with a Dirac monopole of magnetic charge $n$ inserted at the origin)~\cite{borokhov2002}. In the large-$N$ limit, the saddle-point approximation becomes exact; thus in pure QED$_3$ the gauge field can be treated classically and the problem reduces to that of computing the ground-state energy of $N$ free Dirac fermions on $S^2$ in a uniform $2\pi n$ magnetic flux background. In the problem at hand, we also have a complex scalar, but in the large-$N$ limit it can also be treated classically and assumes its saddle-point value. To see this, we simply integrate out all $N$ fermions in Eq.~(\ref{L}), instead of $N-1$ fermions as we did earlier; in the large-$N$ limit, the fermion determinant dominates, and the ground-state energy is given by
\begin{align}\label{E0}
E_0=-\lim_{\beta\rightarrow\infty}\frac{N}{2\beta}\mathrm{Tr\,ln}\,\mathcal{G}^{-1}_{S^2}(A_{\mu,\text{cl}},\phi_\text{cl}),
\end{align}
where $A_{\mu,\text{cl}}(x)$ and $\phi_\text{cl}(x)$ are the saddle-point values, and $\mathcal{G}^{-1}_{S^2}$ is the inverse fermion propagator, but suitably modified on the sphere. Equation~(\ref{E0}) is simply the ground-state energy of $N$ free Dirac fermions in a classical background given by $A_{\mu,\text{cl}}(x)$ and $\phi_\text{cl}(x)$. We expect $A_{\mu,\text{cl}}(x)$ to correspond to uniform magnetic flux on the sphere. Any nonzero saddle-point configuration $\phi_\text{cl}(x)$ for the scalar breaks the gauge symmetry from $U(1)$ to $\mathbb{Z}_2$; since at the critical point the gauge symmetry is unbroken, $\phi_\text{cl}(x)$ must vanish. Thus in the large-$N$ limit the fermions couple only to a classical uniform flux background, as in pure QED$_3$, and the scaling dimension in that limit reduces to that of QED$_3$,
\begin{align}\label{MonopoleQCP}
\Delta_{2\pi}^\text{QCP}=0.265N+\mathcal{O}(1/N^0).
\end{align}

In QED$_3$, the $1/N$ correction---i.e., the second term in Eq.~(\ref{MonopoleQED3})---comes from Gaussian gauge-field fluctuations around the classical flux background~\cite{pufu2014}. In our case, the action for Gaussian fluctuations contains not only gauge-field fluctuations but also fluctuations of the scalar field, which would give an additional contribution to the ground-state energy. Therefore we expect that beginning with the $\mathcal{O}(1/N^0)$ term, the monopole scaling dimension at the $U(1)$-$\mathbb{Z}_2$ quantum critical point (\ref{MonopoleQCP}) will differ from that at the algebraic spin liquid fixed point (\ref{MonopoleQED3}). Leaving the calculation of the $1/N$ correction to $\Delta_{2\pi}^\text{QCP}$ from scalar fluctuations to future work, Eq.~(\ref{MonopoleQCP}) implies that the quantum critical point found is stable against instanton proliferation provided that $N>N_c^\text{inst}$ where
\begin{align}\label{instantonNc}
N_c^\text{inst}=11.32.
\end{align}

\subsection{Critical couplings}

For comparison with RG treatments, such as the $\epsilon$-expansion analysis discussed in the following Sec.~\ref{sec:epsilon}, one can calculate the values $e^{2}_{*}$, $h^{2}_{*}$, and $\lambda_{*}^2$ of the couplings at the RG fixed point corresponding to the critical point found in the large-$N$ limit. To leading order, these critical couplings can be found by demanding that corrections to scaling in the large-$N$ action vanish~\cite{moshe2003}. While this can be done in $d=3$, to facilitate later comparison with the fixed point couplings obtained in the $\epsilon$-expansion we will consider $d=4-\epsilon$ in the limit that $\epsilon\rightarrow0$.

First, $e^{2}_{*}$ can be found by demanding that the coefficient of the $q^{2}$ term in the inverse gauge field propagator $K_{\mu\nu}^{B}$ vanishes exactly. This correlation function has been computed in the limit that $\Lambda\rightarrow\infty$; in order to calculate the critical coupling, we need to compute this same function for finite $\Lambda$. To do this, the following result is needed~\cite{moshe2003}:
\begin{align}\label{FiniteMomInt}
\int_{0}^{\Lambda}dq\frac{q^{d-1}}{q^{2}+4\left|M\right|^{2}} & =  \Lambda^{d-2}\biggl(\frac{1}{d-2}+\frac{\pi}{2\sin\left(\pi d/2\right)}\left|\frac{2M}{\Lambda}\right|^{d-2}\nonumber\\
&-\frac{1}{d-4}\left|\frac{2M}{\Lambda}\right|^{2}+\cdots\biggr).
\end{align}
This result is an expansion in powers of $M/\Lambda$, valid for $2<d<4$. Evaluating the momentum integral in $K_{\mu\nu}^{B}$ with a finite cutoff $\Lambda$, and using Eq.~(\ref{FiniteMomInt}), one obtains the following condition: $1-e^{2}_{*}/(12\pi^2\epsilon)=0$. The same procedure allows $h^{2}_{*}$ to be determined, now by demanding that the coefficient of the $q^{2}$ term in the $\sigma$ inverse propagator vanishes exactly. This leads to the condition: $h^{-2}_{*}-(4\pi^2\epsilon)^{-1}=0$. 
To obtain the critical coupling $\lambda_{*}^2$, we demand the vanishing of corrections to scaling in the gap equation, i.e., the vanishing of the coefficient of $|M|^{2}$ in the gap equation.
Using the finite momentum version of the gap equation, along with Eq.~(\ref{FiniteMomInt}), we obtain $\lambda^{2}_{*}=16h^{4}_{*}/[(4\pi)^2\epsilon]$. By scaling all the coupling constants, $e^2/(4\pi)^2\rightarrow e^2$, $h^2/(4\pi)^2\rightarrow h^2$, $\lambda^2/(4\pi)^2\rightarrow \lambda^2$, and reinserting the prefactor of $1/N$ that was scaled out at the beginning of Sec.~\ref{sec:SaddlePoint}, it follows that the critical couplings in the large-$N$ limit are
\begin{align}\label{largeNFP1}
e_*^2=\frac{3\epsilon}{4N},\hspace{5mm}h_*^2=\frac{\epsilon}{4N},\hspace{5mm}\lambda_*^2=\frac{\epsilon}{N}.
\end{align}

\section{$\epsilon$ expansion}
\label{sec:epsilon}

In the previous section we studied the model described by the Lagrangian in Eq.~(\ref{L}) in fixed spacetime dimension $d=3$, but at the expense of working in the $N\rightarrow\infty$ limit. Alternatively, we can study the theory at finite $N$ but in $d=4-\epsilon$ dimensions, treating $\epsilon$ as a small parameter. This analysis was performed previously in Ref.~\cite{roy2013,roy2016} for general $N$, and in Ref.~\cite{zerf2016} for $N=1$. Owing to some discrepancies in the literature, we revisit this analysis for general $N$. The one-loop RG equations for $N=1$ derived in Ref.~\cite{zerf2016} can be easily generalized to arbitrary $N$ by multiplying all diagrams containing a closed fermion loop by a factor of $N$. Doing so, we obtain the following equations describing the RG flow on the critical hypersurface $m^2=0$~\cite{roy2013,roy2016},
\begin{figure}[t]
\includegraphics[width=0.75\columnwidth]{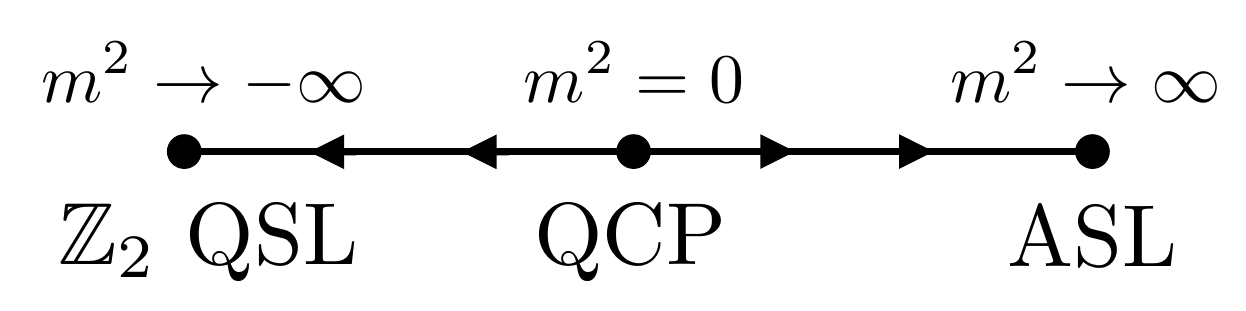}
\caption{RG flows in the $\epsilon$ expansion. The scalar mass squared $m^2$ tunes a transition between the algebraic spin liquid (ASL) and the $\mathbb{Z}_2$ quantum spin liquid (QSL) via a quantum critical point (QCP).}
\label{fig:RG}
\end{figure}
\begin{align}
\beta_{e^2}=&-\epsilon e^2+\frac{4}{3}(N+2)e^4,\label{betae2}\\
\beta_{h^2}=&-\epsilon h^2-12e^2h^2+4(N+2)h^4,\label{betah2}\\
\beta_{\lambda^2}=&-\epsilon\lambda^2+20\lambda^4+8Nh^2\lambda^2-16Nh^4
-48e^2\lambda^2+96e^4,\label{betal2}
\end{align}
where $\beta_{g^2}=dg^2/d\ln\mu$ for $g=e,h,\lambda$, and $\mu$ is a renormalization scale, and we have rescaled the couplings as $g^2/(4\pi)^2\rightarrow g^2$~\footnote{The beta functions appearing in Eqs.~(\ref{betae2})-(\ref{betal2}) can be compared with those in Ref.~\cite{roy2013,roy2016} as follows. Let $g$, $\tilde{e}$, $\tilde{\lambda}$ denote the Yukawa, charge, and scalar couplings in Ref.~\cite{roy2013,roy2016}, and $N_{b}$, $N_{f}$ the number of bosonic and fermionic fields in those references. The relationships to the parameters in this paper are $g^{2}=8h^{2}$, $\tilde{e}^{2}=2e^{2}$, $\tilde{\lambda}=4\lambda^{2}$. In our paper we have $N_{b}=1$, and $N=2N_{f}$; with these identities, it follows that Eqs.~(\ref{betae2}), (\ref{betal2}) are equivalent to Eqs.~(13), (15) in Ref.~\cite{roy2013} and Eq.~(\ref{betah2}) is equivalent to Eq.~(14) of Ref.~\cite{roy2016}.}. The only stable fixed point in a physical range of coupling constants is a fixed point with finite Yukawa coupling and scalar self-interaction~\cite{roy2013,roy2016},
\begin{widetext}
\begin{align}\label{1LFP}
e_*^2=\frac{3\epsilon}{4(N+2)},\hspace{5mm}
h_*^2=\frac{(N+11)\epsilon}{4(N+2)^2},\hspace{5mm}
\lambda_*^2=\frac{76+18N-N^2+\sqrt{N^4+44N^3-2388N^2-4864N-11504}}{40(N+2)^2}\epsilon.
\end{align}
\end{widetext}
This fixed point describes a quantum critical point between the algebraic and $\mathbb{Z}_2$ spin liquids. Indeed, the only relevant direction is the scalar mass squared which flows to either positive or negative infinity (Fig.~\ref{fig:RG}). At the $m^2\rightarrow\infty$ fixed point, the scalar mass term $m^2|\phi|^2$ implies that $\phi$ has dimension 3/2, and all couplings involving the scalar become irrelevant. The Lagrangian reduces to that of QED$_3$, with the one-loop beta function
\begin{align}
\beta_{e^2}^{\text{QED}_3}=-\epsilon e^2+\frac{4}{3}Ne^4,
\end{align}
which admits the stable fixed point $e_{*,\text{QED}_3}^2=3\epsilon/(4N)$~\cite{dipietro2016,dipietro2017} corresponding to the algebraic spin liquid. For $m^2\rightarrow-\infty$, the scalar condenses, Higgsing the gauge field and gapping out the fermions; this is the $\mathbb{Z}_2$ spin liquid. Taking the $N\gg 1$ limit in Eq.~(\ref{1LFP}), the critical couplings
\begin{align}\label{largeNFP}
e_*^2=\frac{3\epsilon}{4N},\hspace{5mm}h_*^2=\frac{\epsilon}{4N},\hspace{5mm}\lambda_*^2=\frac{\epsilon}{N},
\end{align}
reduce to those obtained in Eq.~(\ref{largeNFP1}) of Sec.~\ref{sec:Fluctuations} in fixed dimension $d$ from demanding the vanishing of corrections to scaling (upon setting $d=4-\epsilon$ with $\epsilon\rightarrow 0$), which indicates the large-$N$ and $\epsilon$-expansion approaches are accessing the same fixed point.

The fixed point (\ref{1LFP}) found at one loop only survives above a certain critical number $N_{c}$ of Dirac fermion flavors. This problem bears a certain similarity to scalar QED$_3$ with $N_{b}$ flavors of complex scalars coupled to a fluctuating $U(1)$ gauge field, relevant to the problem of the usual superconducting transition in the presence of dynamical electromagnetism. As is well known from the classic Halperin-Lubensky-Ma analysis~\cite{halperin1974}, at one loop the critical point disappears for $N_b<N_b^c$ with $N_b^c=182.95$. The destruction of the critical point proceeds via its annihilation at $N=N_b^c$ with another, bicritical point, and subsequent disappearance into the complex plane for $N_b<N_b^c$. (This is an instance of the generic fixed-point annihilation scenario described in Ref.~\cite{kaplan2009}, and argued also to be relevant for dynamical mass generation in {\it fermionic} QED$_3$ at low fermion flavor number~\cite{herbut2016}.) In other words, the critical point disappears when the two values $\lambda_{*,\pm}^2$ for the scalar self-interaction at the two fixed points---which are given by expressions similar to $\lambda_*^2$ in Eq.~(\ref{1LFP}), but with $\pm$ in front of the square root---become equal to each other. Thus $N_b^c$ is given by the condition that the argument of the square root vanishes. However, the situation is different in the problem at hand. 
For scalar QED$_{3}$ both roots $\lambda_{*,\pm}^2$ remain positive all the way down to $N_b=N_b^c$. In our case, however, as $N$ decreases from infinity the stable (critical) root $\lambda_{*}^{2}$ given in Eq.~(\ref{1LFP})  becomes negative {\it before} annihilating with another root.
This other root, given by Eq.~(\ref{1LFP}) but with a minus sign in front of the square root, is unphysical for all $N$. For negative $\lambda^2$ the Landau-Ginzburg action should be supplemented by a $\sim|\phi|^6$ term, and the transition becomes first order. Therefore $N_c$ is given not by the condition that the argument of the square root vanishes, but by the condition $\lambda^{2}_{*}=0$. This gives the value
\begin{align}\label{Nc}
N_c=34.90,
\end{align}
which is a more stringent condition than that of stability against instanton proliferation (\ref{instantonNc}). Below $N=N_c$, the two (unphysical) roots ultimately annihilate at the smaller value $N_c'=33.04$~\cite{roy2016}.

Finally, for comparison, we discuss the distinction between the present model for an algebraic spin liquid-$\mathbb{Z}_{2}$ spin liquid transition and the model~\cite{janssen2017} of an algebraic spin liquid-chiral spin liquid transition. In the latter model a real scalar field is coupled to the time-reversal symmetry-breaking fermion mass term, rather than the fermion pairing term, while the gauge field couples only to the fermions. Unlike a real scalar, which is electrically neutral and does not directly couple to the gauge field, the complex scalar in our case is charged and 
does directly couple to the gauge field. This coupling tends to destabilize an otherwise continuous transition due to the fluctuation-induced first-order transition mechanism ~\cite{halperin1974}. This explains in part why in Ref.~\cite{janssen2017} the phase transition is continuous all the way down to $N=2$. In the present model, adding $N$ massless fermions partially remedies the destabilization induced by the complex scalar coupling, at least for sufficiently large $N$, 
as the analysis in this section shows. For small $N$ the problem becomes closer to the pure scalar $\mathrm{QED}_{3}$ problem which exhibits a fluctuation-induced first-order transition to lowest order in the $\epsilon$ expansion.

\subsection{Critical exponents}

\begin{figure}[t]
\includegraphics[width=0.75\columnwidth]{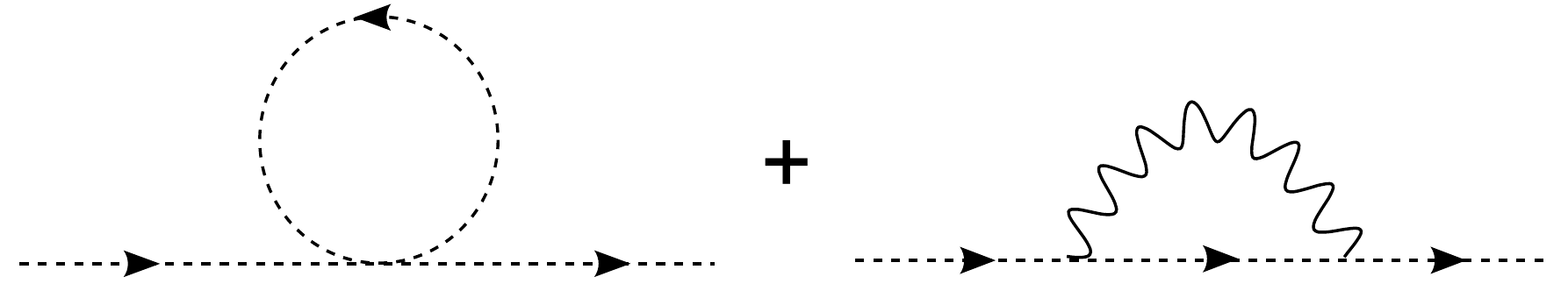}
\caption{The two diagrams contributing to scalar mass renormalization in a general $\xi$-gauge (dotted line: scalar, wiggly line: gauge field).}
\label{fig:Zm2}
\end{figure}

Assuming $N>N_c$, one has a novel universality class of fermionic quantum critical points with non-Gaussian critical exponents. The correlation length exponent $\nu$ can be determined in the $\epsilon$ expansion by deriving the RG flow equation for the scalar mass squared,
\begin{align}
\frac{dm^2}{d\ln\mu}=(-2+\gamma_\phi-\gamma_{m^2})m^2,
\end{align}
where $\gamma_\phi=d\ln Z_\phi/d\ln \mu$ and $\gamma_{m^2}=d\ln Z_{m^2}/d\ln\mu$, and $Z_\phi$ and $Z_{m^2}$ are the renormalization constants appearing in the renormalized Lagrangian
\begin{align}\label{Lren}
\mathcal{L}=&\sum_{a=1}^{N}iZ_\psi\bar{\psi}_a(\slashed{\partial}+ie\mu^{\epsilon/2}\slashed{A})\psi_a+\frac{1}{4}Z_AF_{\mu\nu}^2+\frac{1}{2\xi}(\partial_\mu A_\mu)^2\nn\\
{}&+Z_\phi|(\partial_\mu+2ie\mu^{\epsilon/2}A_\mu)\phi|^2
+Z_{m^2}m^2\mu^2|\phi|^2\nn\\
{}&+Z_{\lambda^2}\lambda^2\mu^\epsilon|\phi|^4+Z_hh\mu^{\epsilon/2}\sum_{a=1}^{N}\left(\phi^*\psi_a^Ti\sigma_2\psi_a+\text{H.c.}\right).
\end{align}
The correlation length exponent is the inverse of the RG eigenvalue of the scalar mass operator at the fixed point:
\begin{align}\label{nuinv}
\nu^{-1}=2-\gamma_\phi^*+\gamma_{m^2}^*.
\end{align}
The renormalization constant $Z_\phi$ was computed in Eq.~(A6) of Ref.~\cite{zerf2016} for $N=1$; for general $N$ we simply need to multiply the coefficient of the term $\propto h^2$ by $N$, since it corresponds to a scalar self-energy diagram with a closed fermion loop:
\begin{align}
Z_\phi=1-\frac{4Nh^2}{\epsilon}+\frac{8e^2(3-\xi)}{\epsilon},
\end{align}
in a general $\xi$-gauge, which provides an additional check on the calculation. In the minimal subtraction scheme, only two diagrams contribute at one loop to the scalar mass renormalization constant $Z_{m^2}$ (Fig.~\ref{fig:Zm2}), and we obtain
\begin{align}
Z_{m^2}=1+\frac{8\lambda^2}{\epsilon}-\frac{8e^2\xi}{\epsilon}.
\end{align}

The logarithmic derivatives $\gamma_\phi$ and $\gamma_{m^2}$ are then 
\begin{align}
\gamma_\phi&=\frac{1}{Z_\phi}\left(-\frac{4N}{\epsilon}\beta_{h^2}+\frac{8(3-\xi)}{\epsilon}\beta_{e^2}-\frac{8e^2}{\epsilon}\beta_\xi\right),\\
\gamma_{m^2}&=\frac{1}{Z_{m^2}}\left(\frac{8}{\epsilon}\beta_{\lambda^2}-\frac{8\xi}{\epsilon}\beta_{e^2}-\frac{8e^2}{\epsilon}\beta_\xi\right),
\end{align}
where $\beta_\xi=d\xi/d\ln\mu$. Inserting those expressions into Eq.~(\ref{nuinv}), and expanding to leading order in $\epsilon$, we find that the gauge-dependent terms cancel, and
\begin{align}
\nu^{-1}=2-4Nh_*^2+24e_*^2-8\lambda_*^2,
\end{align}
using Eqs.~(\ref{betae2})-(\ref{betal2}). This expression agrees with the corresponding one in Eq.~(16) of Ref.~\cite{roy2013} (where $N_{b}=1$ in our case). Substituting the critical couplings (\ref{1LFP}), we obtain
\begin{widetext}
\begin{align}
\nu=\frac{1}{2}+\frac{4N^2-17N-104+\sqrt{N^4+44N^3-2388N^2-4864N-11504}}{20(N+2)^2}\epsilon+\mathcal{O}(\epsilon^2).
\end{align}
\end{widetext}

One can also study the scaling dimension of fermion mass operators. Here there is an added subtlety owing to the different spinor structures in $d=4$ and $d=3$ dimensions~\cite{dipietro2016,dipietro2017}. For the rest of this section we consider an even number $N\equiv 2N_f$ of two-component Dirac fermions $\psi_a$. In the $\epsilon$ expansion one is really attempting to dimensionally continue $N_f$ four-component Dirac spinors $\Psi_a$ in the $d=4$ Lorentz algebra to $d<4$. In a theory with massless fermions this is equivalent to the two-component formulation if the two Weyl components of $\Psi_a$ are interpreted as distinct two-component Dirac spinors in the $d=3$ Lorentz algebra,
\begin{align}
\Psi_a=\left(\begin{array}{c}
\psi_a \\
\psi_{a+N_f}
\end{array}\right),\hspace{5mm}a=1,\ldots,N_f.
\end{align}
However, adding a mass term $iM\sum_{a=1}^{N_f}\bar{\Psi}_a\Psi_a$ in $d=4$ really corresponds to
\begin{align}\label{fermionmassOP}
iM\sum_{a=1}^{N_f}\bar{\psi}_a\psi_{a+N_f}+\mathrm{H.c.}
\end{align}
in $d=3$~\cite{dipietro2016,dipietro2017}.

We can calculate the scaling dimension $\Delta$ of the fermion mass operator in Eq.~(\ref{fermionmassOP}) by adding it to the Lagrangian, multiplied by a renormalization constant $Z_M$, and deriving the flow equation for $M$:
\begin{align}
\frac{dM}{d\ln\mu}=(-1+\gamma_\psi-\gamma_M)M,
\end{align}
where $\gamma_\psi=d\ln Z_\psi/d\ln\mu$ and $\gamma_M=d\ln Z_M/d\ln\mu$. The scaling dimension $\Delta$ is given by
\begin{align}
d-\Delta=1-\gamma_\psi^*+\gamma_M^*,
\end{align}
where the right-hand side is evaluated at the fixed point (\ref{1LFP}), with $\gamma_\psi^*$ the fermion anomalous dimension. While $\gamma_\psi^*$ itself is not meaningful, since $\psi$ is not a gauge invariant operator, we expect that all gauge dependence should cancel in $\Delta$ because the fermion mass term is a gauge invariant operator.

\begin{figure}[t]
\includegraphics[width=0.75\columnwidth]{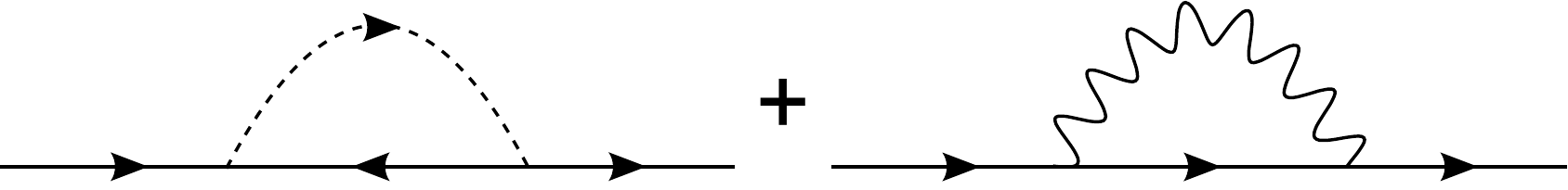}
\caption{The two diagrams contributing to fermion mass renormalization in a general $\xi$-gauge (solid line: fermion, dotted line: scalar, wiggly line: gauge field).}
\label{fig:Zmpsi}
\end{figure}
Two diagrams contribute to the fermion mass renormalization (Fig.~\ref{fig:Zmpsi}). At one-loop order, we obtain
\begin{align}
Z_M=1+\frac{8h^2}{\epsilon}-\frac{2(3+\xi)e^2}{\epsilon},
\end{align}
and $Z_\psi$ was calculated in Ref.~\cite{zerf2016},
\begin{align}
Z_\psi=1-\frac{4h^2}{\epsilon}-\frac{2e^2\xi}{\epsilon},
\end{align}
noting that general $N$ does not affect the fermion self-energy at one-loop order. Proceeding as in the above, we find
\begin{align}\label{FermionMass}
d-\Delta=1-12h_*^2+6e_*^2,
\end{align}
where we see that all $\xi$ dependence cancels, as expected. At the algebraic spin liquid fixed point, where $h_*^2=0$ and $e_*^2=3\epsilon/(8N_f)$, we recover the one-loop result previously derived in Ref.~\cite{dipietro2016,dipietro2017}. At the $U(1)$-$\mathbb{Z}_2$ quantum critical point, substituting in the fixed point couplings in Eq.~(\ref{1LFP}) gives
\begin{align}
d-\Delta=1+\frac{3(N_f-8)}{4(N_f+1)^2}\epsilon+\mathcal{O}(\epsilon^2).
\end{align}

\subsection{Three-loop correction to critical number of fermion flavors}

The quantum critical point discussed above at one-loop order only exists for an anomalously large number $N>34.90$ of fermion flavors, which is similar to scalar QED for which $N_b^c=182.95$. In scalar QED, however, this is possibly an artefact of the one-loop calculation, as it is known that corrections at higher loop order, which are potentially large in physical dimensions $\epsilon=1$, can dramatically change the critical value of $N_b$~\cite{IgorBook}. To investigate whether this also happens in the present problem, 
we compute corrections to $N_c$ up to $\mathcal{O}(\epsilon^3)$.

To perform this analysis we must first compute the RG beta functions at three-loop order. To do this we use the minimal subtraction scheme with dimensional regularization in $d=4-\epsilon$ spacetime dimensions, following the method described in Ref.~\cite{zerf2016} but generalized to $N$ fermion flavors. The full expressions for the three-loop order beta functions are rather lengthy and are given in Appendix~\ref{sec:Appendix_BetaFunc}. Here we give the beta functions at two-loop order only:
\begin{widetext}
\begin{align}
\beta_{e^2}=&-\epsilon e^2+\frac{4}{3}(N+2)e^4+4(N+32)e^6-8Nh^2e^4,\label{betae22L}\\
\beta_{h^2}=&-\epsilon h^2-12e^2h^2+4(N+2)h^4+8h^2\lambda^4-\frac{2}{3}(2N+349)h^2e^4-64h^4\lambda^2+4(5N+88)h^4e^2-8(6N-7)h^6,\label{betah22L}\\
\beta_{\lambda^2}=&-\epsilon\lambda^2+20\lambda^4+8Nh^2\lambda^2-16Nh^4
-48e^2\lambda^2+96e^4+16N\lambda^2h^4+\frac{16}{3}(5N+316)\lambda^2e^4+40N\lambda^2h^2e^2\nn\\
{}&-80N\lambda^4h^2
+448\lambda^4e^2-240\lambda^6+256Nh^2e^4+64Nh^4e^2
+256Nh^6-\frac{512}{3}(N+26)e^6.\label{betal22L}
\end{align}
\end{widetext}

To determine corrections to $N_{c}$, we follow the method discussed in Ref.~\cite{IgorBook} for scalar QED, but suitably adapted to the problem at hand. As already mentioned above, in scalar QED the critical flavor number $N_b^c$ is determined from the condition that the critical and bicritical points merge into a single fixed point. At the one-loop level this produces the value $N_b^c=182.95$ quoted above; to obtain $\mathcal{O}(\epsilon)$ corrections to this value one demands that the fixed points of the two-loop order beta functions likewise merge into a single fixed point. As explained above, in our problem $N_{c}$ is obtained from the condition $\lambda^{2}_{*}=0$, which is more stringent than the condition of fixed-point annihilation. Likewise, to obtain 
$\mathcal{O}(\epsilon)$ corrections to Eq.~(\ref{Nc}) we require that the fixed-point value $\lambda^2_{*}$ vanishes at two-loop order, and similarly to determine the $\mathcal{O}(\epsilon^{2})$ corrections we require that $\lambda^{2}_{*}$ vanishes at three-loop order. We thus look for fixed-point couplings of the form
\begin{align}
e_*^2&=\epsilon(x_0+x_{1}\epsilon+x_{2}\epsilon^2),\label{eq1}\\
h_*^2&=\epsilon(y_0+y_{1}\epsilon+y_{2}\epsilon^2),\\
\lambda_*^2&=0,\\
N_c&=N_c^0+z_{1}\epsilon+z_{2}\epsilon^{2},\label{eq4}
\end{align}
where $N_c^0=34.90$ is the one-loop value (\ref{Nc}), and
\begin{align}
x_0=\frac{3}{4(N_c^0+2)},\hspace{5mm}
y_0=\frac{N_c^0+11}{4(N_c^0+2)^2},
\end{align}
are the one-loop fixed-point couplings exactly at $N=N_c^0$, i.e., the tricritical point below which the transition becomes first order. We substitute Eqs.~(\ref{eq1})-(\ref{eq4}) into the two-loop order beta functions Eqs.~(\ref{betae22L})-(\ref{betal22L}), set the beta functions equal to zero, and expand the resulting equations to lowest nontrivial order in $\epsilon$. This gives a system of three linear equations for the three unknowns $x_{1}, y_{1},$ and $z_{1}$, which are easily solved numerically to give $z_{1}=-76.67$. 
Inserting Eqs.~(\ref{eq1})-(\ref{eq4}) into the three-loop order beta functions (see Appendix \ref{sec:Appendix_BetaFunc}), and requiring these to vanish to lowest nontrivial order in $\epsilon$ then gives a system of simultaneous equations for $x_{2}, y_{2},$ and $z_{2}$, the solution of which yields $z_{2}=3.993$. Therefore, the critical number of fermion flavors is
\begin{align}
N_c=N_c^0\left(1-2.197\epsilon+0.1144\epsilon^{2}+\mathcal{O}(\epsilon^3)\right).
\end{align}
Thus, as in scalar QED~\cite{IgorBook}, the leading correction to the critical flavor number $N_c$ is negative; the subleading correction is positive but with a coefficient that is an order of magnitude smaller. A naive extrapolation to physical dimensions ($\epsilon=1$) produces a negative value of $N_c$, which suggests that the $U(1)$-$\mathbb{Z}_2$ spin liquid transition remains continuous for any value of $N$. Of course, such an extrapolation is uncontrolled, and only an exact determination of the beta function (or other nonperturbative methods) can determine the precise value of $N_c$.

\section{Conclusion}
\label{sec:conclusion}

We have proposed a field theory description of a quantum phase transition between a gapless algebraic $U(1)$ spin liquid with $N$ flavors of fermionic Dirac spinons and a gapped $\mathbb{Z}_2$ spin liquid, where the gap is opened by spinon pair condensation and ensuing Higgsing of the emergent $U(1)$ gauge field. Using large-$N$ and $\epsilon$-expansion techniques, we showed that the transition---described by a strongly coupled (2+1)D conformal gauge theory with non-Gaussian exponents---is continuous for sufficiently large values of $N$. Conversely, we also showed that one expects the critical point to disappear, either by flowing to a different universality class, or by becoming a first-order transition, for $N$ less than some critical value $N_c$. Different approaches yielded different estimates of $N_c$. In the large-$N$ limit a natural criterion for $N_c$ is the value of $N$ below which monopole operators, corresponding to instanton events in spacetime driving spinon confinement, become relevant in the RG sense. This produced the estimate $N_c^\text{inst}=11.32$, essentially the value below which the algebraic spin liquid itself is destroyed by instantons. 

In the $\epsilon$ expansion we found $N_c=34.90$ at one-loop order, below which the coefficient of the quartic term for the spinon pair order parameter in the Landau-Ginzburg-Wilson action becomes negative and the transition becomes first order. Unfortunately, both values are greater than physically relevant values in currently known spin liquid candidates, such as $N=4$ in the proposed algebraic spin liquid ground state of the spin-1/2 Heisenberg model on the kagom\'e lattice~\cite{ran2007,hermele2008}, relevant for herbertsmithite~\cite{han2012,fu2015}. The following caveats, however, are in order. First, we showed that when extrapolated to 2+1 dimensions, corrections up to three-loop order in the $\epsilon$ expansion make $N_c$ formally negative, leading to the possibility that in physical dimensions the transition might remain continuous in experimentally accessible systems. Second, values of $N$ several times greater than $N=4$ are potentially realizable in cold atom systems~\cite{gorshkov2010,desalvo2010}. Third, and as already mentioned implicitly, according to the large-$N$ result at next-to-leading order [Eq.~(\ref{MonopoleQED3})] the $N=4$ algebraic spin liquid itself is naively unstable to spinon confinement---although taking into account the space/point group quantum numbers carried by the monopole operators in a given microscopic spin model can increase the magnetic charge of the smallest symmetry-allowed monopole operator and reduce the critical value of $N$. At any rate, the question of $N_c$ needs further investigation not only at transitions between different spin liquid states but also in the algebraic spin liquid phase itself.

To test our predictions, nonperturbative methods such as numerical studies of microscopic spin models are of paramount importance. For instance, different density-matrix renormalization group studies suggest that the ground state of the kagom\'e lattice Heisenberg antiferromagnet is either a gapped $\mathbb{Z}_2$ spin liquid~\cite{yan2011,jiang2012} or the algebraic spin liquid mentioned above~\cite{he2017,zhu2018} (which would imply that in reality $N_c<4$ at the latter fixed point). The difficulty in pinning down the nature of the ground state in this model suggests that the two candidate states are close in energy, and thus that the transition discussed here might be possibly accessed and studied by adding and tuning various symmetry-preserving perturbations. If such a transition can be found in a microscopic $SU(2)$-symmetric spin model, or more generally a model with $SU(k)$ spins, a numerical determination of the order of the transition as a function of $k$ could be compared with our leading-order prediction of a transition that is first order for $N<N_c$ and continuous for $N>N_c$.

Besides the search for $U(1)$-$\mathbb{Z}_2$ spin liquid transitions in numerical studies of microscopic spin models, other avenues present themselves for future work. First, critical exponents such as $\nu$ and the scaling dimension of other gauge-invariant operators, including the fermion mass term, can be calculated beyond leading order in the $1/N$ expansion in fixed $d=3$ dimensions. Already at order $1/N$ we expect mixing between the gauge and matter sectors, which did not occur in the strict $N\rightarrow\infty$ limit. Combined with unitarity bounds in conformal field theory~\cite{rychkov2016,simmons-duffin2016}, this may give alternate restrictions on the allowed values of $N$~\cite{pufu2014}. As already mentioned, using the methods of Ref.~\cite{pufu2014} one can in principle also calculate next-to-leading order corrections to the scaling dimension of monopole operators at the quantum critical point, which would improve the estimate in Eq.~(\ref{instantonNc}). 

Furthermore, as done in Ref.~\cite{grover2014b}, one can also potentially derive other estimates of $N_c$ by exploiting the $F$-theorem. According to this theorem, the universal part $F$ of the free energy of a (2+1)D conformal field theory on the three-sphere $S^3$ decreases monotonically along RG flows~\cite{casini2012}. In the present context, this implies that $F_\text{QCP}\geq F_\text{ASL}$ and $F_\text{QCP}\geq F_{\mathbb{Z}_2}$ in obvious notation (see Fig.~\ref{fig:RG}). For a gapped topological phase such as the $\mathbb{Z}_2$ spin liquid, $F$ corresponds to the topological entanglement entropy~\cite{levin2006,kitaev2006b}, which is known to be $F_{\mathbb{Z}_2}=\ln 2$ for $\mathbb{Z}_2$ topological order. In principle, $F_\text{QCP}$ could be computed in the large-$N$ limit (see, e.g., Ref.~\cite{klebanov2012}). For a coarse estimate, we expect to be able to approximate $F_\text{QCP}\approx F_\text{ASL}$ in the large-$N$ limit since the number $N$ of fermions greatly exceeds that (one) of scalars. Using $F_\text{ASL}\approx 0.22N+\frac{1}{2}\ln(\pi N/8)$ valid in the large-$N$ limit~\cite{klebanov2012}, setting $F_\text{QCP}\geq F_{\mathbb{Z}_2}$ implies $N\geq 2.88$, which is a much less stringent bound than those already derived. Finally, numerical studies of the field theory (\ref{L}) using the conformal bootstrap approach~\cite{simmons-duffin2016} may help in achieving a nonperturbative determination of the critical number $N_c$ of fermion flavors.

\acknowledgements

We thank I. Affleck, K. Wamer, and W. Witczak-Krempa for helpful correspondence. RB was supported by the Theoretical Physics Institute at the University of Alberta. AR was supported by NSERC and the Government of Alberta. JM was supported by NSERC grant \#RGPIN-2014-4608, the CRC Program, CIFAR, and the University of Alberta.

\appendix
\numberwithin{equation}{section}
\numberwithin{figure}{section}
\section{Large-$N$ inverse propagators}
\label{sec:app_Fluctuations}
Here we present a brief derivation of the results presented in Eqs.~(\ref{eq:B_Response}-\ref{eq:Sigma_Response}). The gauge-field action is given by $\frac{1}{4}\mathop{\mathrm{Tr}}\mathcal{G}_{\mathrm{cl}}X_{B}\mathcal{G}_{\mathrm{cl}}X_{B}$.
In momentum space the propagator is 
\begin{equation}
\left.\mathcal{G}\right|_{\phi=\phi_{\mathrm{cl}},A_{\mu}=A_{\mu,\mathrm{cl}}}=\frac{1}{k^{2}+4\left|M\right|^{2}}\left(\begin{array}{cc}
2M & \slashed{k}\\
\slashed{k} & -2M^{*}
\end{array}\right).
\end{equation}
At the phase transition point, $M=0$, and so we henceforth consider this case only. Converting to momentum space, it follows that the gauge-field action is given by
\begin{align}
\frac{1}{2}\int\frac{d^dq}{(2\pi)^d}B_{\mu}(q)K^{B}_{\mu\nu}(q)B_{\nu}(-q),
\end{align} 
where the inverse propagator is
\begin{equation}
K^{B}_{\mu\nu}\left(q\right)= \frac{e^{2}}{2}\mathrm{Tr}\,\mathcal{G}\left(k+q\right)\left(\begin{array}{cc}
0 & \gamma_{\mu}\\
-\gamma_{\mu} & 0
\end{array}\right)\mathcal{G}\left(k\right)\left(\begin{array}{cc}
0 & \gamma_{\nu}\\
-\gamma_{\nu} & 0
\end{array}\right).
\end{equation}
After taking the trace over Nambu indices, the integral that remains is given by 
\begin{equation}
K^{B}_{\mu\nu}\left(q\right)=e^{2}\mathrm{tr}\int\frac{d^{d}k}{\left(2\pi\right)^{d}}\frac{\left(\slashed{k}+\slashed{q}\right)\gamma_{\mu}\slashed{k}\gamma_{\nu}}{\left(k+q\right)^{2}k^{2}}.
\end{equation}
To compute the momentum integral we introduce the Feynman parameter $x$, defined by
\begin{equation}\label{eq:FeynPar}
\frac{1}{\left(k+q\right)^{2}k^{2}} = \int_{0}^{1}\frac{dx}{\left[\left(k+xq\right)^{2}+\Delta\right]^{2}},
\end{equation}
where $\Delta=q^{2}x\left(1-x\right)$. In $d=3$ dimensions the trace algebra can be performed, and the results are
\begin{eqnarray}
\mathrm{tr}\gamma_{\mu}\gamma_{\nu} & = & 2\delta_{\mu\nu},\\
\mathrm{tr}\gamma_{\alpha}\gamma_{\mu}\gamma_{\beta}\gamma_{\nu} & = & 2\left[\delta_{\alpha\mu}\delta_{\beta\nu}+\delta_{\mu\beta}\delta_{\alpha\nu}-\delta_{\alpha\beta}\delta_{\mu\nu}\right].
\end{eqnarray}
After performing the trace algebra, and simplifying, the inverse propagator becomes
\begin{eqnarray}
K^{B}_{\mu\nu}\left(q\right)&=&2e^{2}\int_{0}^{1}dx\int\frac{d^{d}L}{\left(2\pi\right)^{d}}\biggl[L^{2}\left(\frac{2}{d}-1\right)\delta_{\mu\nu}\nonumber\\
&&-\Delta\left(\frac{2q_{\mu}q_{\nu}}{q^2}-\delta_{\mu\nu}\right)\biggr]\frac{1}{\left[L^{2}+\Delta\right]^{2}}.
\end{eqnarray}
Using standard results~\cite{ZJ}, the inverse propagator is then 
\begin{eqnarray}
 K^{B}_{\mu\nu}\left(q\right)&=&\frac{4e^{2}\left(1-d/2\right)}{\left(4\pi\right)^{d/2}}\left(\delta_{\mu\nu}-\frac{q_{\mu}q_{\nu}}{q^{2}}\right)\nonumber\\
 &&\times\Gamma\left(1-d/2\right)\int_{0}^{1}dx\left(\frac{1}{\Delta}\right)^{1-d/2}.\end{eqnarray}
The integration over $x$ is easily computed~\cite{Whittaker_WatsonBook}, with the final result given by
\begin{eqnarray}
K^{B}_{\mu\nu}\left(q\right)
 & = &e^{2}\left(\delta_{\mu\nu}-\frac{q_{\mu}q_{\nu}}{q^{2}}\right)\left(q^{2}\right)^{d/2-1}\nonumber\\
 && \times\frac{\pi^{(3-d)/2}\left(2-d\right)}{4^{d-1}\Gamma\left((d+1)/2\right)\sin\left(\pi d/2\right)}.
\end{eqnarray}
Evaluating this expression in $d=3$ dimensions, we obtain the result in Eq.~(\ref{eq:B_Response}) of the main text:
\begin{equation}
K^{B}_{\mu\nu}\left(q\right)=\frac{1}{16}e^{2}\left(\delta_{\mu\nu}-\frac{q_{\mu}q_{\nu}}{q^{2}}\right)\left|q\right|.
\end{equation}
This result agrees with the large-$N$ analysis for $\mathrm{QED}_{3}$ with massless 
fermions~\cite{Bhattacharyya_1987}, but is $1/2$ of the result in Ref.~\cite{kim1999}, which is due to the fact that they have four-component
spinors as opposed to the two-component spinors used in this work.

The calculation of the scalar field diagram proceeds along a similar analysis. The diagram is $\frac{1}{4}\mathop{\mathrm{Tr}}\mathcal{G}_{\mathrm{cl}}X_{\sigma}\mathcal{G}_{\mathrm{cl}}X_{\sigma}$.
Converting to momentum space, the action is
\begin{align}
\int\frac{d^dq}{(2\pi)^d}\sigma^{*}(q)K_{\sigma^{*},\sigma}(q)\sigma(q),
\end{align}
where the inverse propagator is
\begin{equation}
K_{\sigma^{*},\sigma}\left(q\right)=\frac{1}{2}\mathrm{Tr}\,\mathcal{G}\left(k+q\right)\left(\begin{array}{cc}
2 & 0\\
0 & 0
\end{array}\right)\mathcal{G}\left(k\right)\left(\begin{array}{cc}
0 & 0\\
0 & -2
\end{array}\right).
\end{equation}
After taking the trace over Nambu indices, the integral that remains is given by 
\begin{equation}
K_{\sigma^{*},\sigma}\left(q\right)=-2\mathrm{tr}\int\frac{d^{d}k}{\left(2\pi\right)^{d}}\frac{\left(\slashed{k}+\slashed{q}\right)\slashed{k}}{\left(k+q\right)^{2}k^{2}}.
\end{equation}
Using the Feynman parameter relation defined in Eq.~({\ref{eq:FeynPar}), along with the trace algebra relations, and then simplifying, we obtain 
\begin{equation}
K_{\sigma^{*},\sigma}\left(q\right)=-4\int_{0}^{1}dx\int\frac{d^{d}L}{\left(2\pi\right)^{d}}\frac{L^{2}-\Delta}{\left[L^{2}+\Delta\right]^{2}}.
\end{equation}
Computing the $L$ integration using standard results~\cite{ZJ} then gives
\begin{equation}
K_{\sigma^{*},\sigma}\left(q\right)=-\frac{4}{\left(4\pi\right)^{d/2}}(d-1)\Gamma\left(1-d/2\right)\int_{0}^{1}dx\left(\frac{1}{\Delta}\right)^{1-d/2}.
\end{equation}
Finally the $x$ integration can be performed~\cite{Whittaker_WatsonBook}, which produces 
\begin{equation}
K_{\sigma^{*},\sigma}\left(q\right)=-\frac{\pi^{(3-d)/2}\left(d-1\right)}{4^{d-3/2}\Gamma\left((d+1)/2\right)\sin\left(\pi d/2\right)}\left(q^{2}\right)^{d/2-1}.
\end{equation}
Evaluating this expression in $d=3$ dimensions, we obtain the result in Eq.~(\ref{eq:Sigma_Response}) of the main text:
\begin{equation}
K_{\sigma^{*},\sigma}\left(q\right)=\frac{1}{4}\left|q\right|.
\end{equation}

\section{Three-loop order beta functions at finite $N$}
\label{sec:Appendix_BetaFunc}
Here we write down the expressions for the three-loop order beta functions at finite $N$:
\begin{widetext}
\begin{align}
\beta_{e^{2}}&=-\epsilon e^{2}+\frac{1}{3}(8+4N)e^{4}+(4N+128)e^{6}-8Ne^{4}h^{2}+\frac{1}{9}(13568-910N-22N^2)e^{8}-716Ne^{6}h^{2}\nonumber\\
&\quad+256e^{6}\lambda^{2}+(112N+12N^{2})e^{4}h^{4}-32e^{4}\lambda^{4},\\
\beta_{h^{2}}&=-\epsilon h^{2}-12e^{2}h^{2}+(8+4N)h^{4}-\frac{1}{3}(698+4N)e^{4}h^{2}+(352+20N)e^{2}h^{4}+(56-48N)h^{6}-64h^{4}\lambda^{2}+8h^{2}\lambda^{4}\nonumber\\
&\quad +\left[\frac{1}{27}(177446+11420N+260N^{2})-768\zeta(3)-96N\zeta(3)\right]e^{6}h^{2}+(96\zeta(3)-1566-624N\zeta(3)+641N-12N^2)e^{4}h^{4}\nonumber\\
&\quad -448e^{4}h^{2}\lambda^{2}+(32-2040N-96N^{2})e^{2}h^{6}-1984e^{2}h^{4}\lambda^{2}+320e^{2}h^{2}\lambda^{4}\nonumber\\
&\quad +(384\zeta(3)-1816+192N\zeta(3)+60N+104N^{2})h^{8}+(1152+480N)h^{6}\lambda^{2}+(360-60N)h^{4}\lambda^{4}-40h^{2}\lambda^{6},\\
\beta_{\lambda^{2}}&= -\epsilon\lambda^{2}+96e^{4}-48e^{2}\lambda^{2}-16Nh^{4}+8Nh^{2}\lambda^{2}+20\lambda^{4}-\frac{1}{3}(13312+512N)e^{6}+256Ne^{4}h^{2}+\frac{1}{3}(5056+80N)e^{4}\lambda^{2}\nonumber\\
&\quad +64Ne^{2}h^{4}+40Ne^{2}h^{2}\lambda^{2}+448e^{2}\lambda^{4}+256Nh^{6}+16Nh^{4}\lambda^{2}-80Nh^{2}\lambda^{4}-240\lambda^{6}\nonumber\\
&\quad+\left[73728\zeta(3)+3072N\zeta(3)-\frac{1}{9}\left(653056+17056N+640N^{2}\right)\right]e^{8}+(19680N+7680N\zeta(3)+512N^{2})e^{6}h^{2}\nonumber\\
&\quad+\left[18432\zeta(3)-384N\zeta(3)+\frac{1}{27}\left(4781696+71768N+560N^{2}\right)\right]e^{6}\lambda^{2}+(7992N-19968N\zeta(3)-688N^{2})e^{4}h^{4} \nonumber\\
&\quad -(17870N+1632N\zeta(3)+24N^{2})e^{4}h^{2}\lambda^{2}-(63648+36864\zeta(3)+616N)e^{4}\lambda^{4}+(12992N-6144N\zeta(3))e^{2}h^{6}\nonumber\\
&\quad +(18048N\zeta(3)-16504N-192N^{2})e^{2}h^{4}\lambda^{2}+(1988N-3264N\zeta(3))e^{2}h^{2}\lambda^{4}-3456e^{2}\lambda^{6}\nonumber\\
&\quad +(1696N-3072N\zeta(3)-2464N^{2})h^{8}+(2048N^{2}-10712N-2688N\zeta(3))h^{6}\lambda^{2}\nonumber\\
&\quad +(4072N+3264N\zeta(3)-240N^{2})h^{4}\lambda^{4}+904Nh^{2}\lambda^{6} +(4936+3072\zeta(3))\lambda^{8}.
\end{align}
For the $N=1$ case the above results are in agreement with Ref.~\cite{zerf2016}.
\end{widetext}

\bibliography{U1Z2QSL}

\end{document}